\documentclass[a4paper]{article}
\usepackage{spconf,amsmath,graphicx,cite}
\usepackage{amsthm}
\usepackage{epstopdf}
\usepackage[titletoc]{appendix}

\title{\vspace*{-0.4cm}Optimal Multiuser Scheduling Schemes for Simultaneous Wireless Information and Power Transfer}

\name{Maryna Chynonova, Rania Morsi, Derrick Wing Kwan Ng, and Robert Schober\thanks{This work was supported in part by the AvH Professorship Program of the Alexander von Humboldt Foundation.  Derrick Wing Kwan Ng and Robert Schober are also with the University of British Columbia, Vancouver, Canada.}\vspace*{-3mm}}
\address{Institute for Digital Communications\\ Friedrich-Alexander-University Erlangen-N{\"u}rnberg, Germany\vspace*{-3mm} }
\usepackage{acronym}

\newcommand{\Un}{\text{U}}
\newcommand{\Sa}{\mathcal{S}_{\text{a}}}

\newcommand{\argord}{\operatornamewithlimits{argorder}}
\newtheorem{Remark}{Remark}
\newtheorem{Thm}{Theorem}
\newtheorem{Prob}{Problem}

\DeclareMathOperator{\mino}{minimize}
\begin{document}

 \textheight 9.9in
 \voffset -0.0in
\maketitle

\begin{abstract}

In this paper, we study the downlink multiuser scheduling problem for systems with simultaneous wireless information and power transfer (SWIPT).   We design optimal scheduling algorithms that maximize the long-term average  system throughput under different fairness requirements, such as  proportional fairness and equal throughput fairness. In particular, the algorithm designs are formulated as  non-convex optimization problems which take into account the minimum required  average sum harvested energy in the system.   The problems are solved by using convex optimization techniques and the proposed optimization framework reveals the tradeoff between the long-term average system throughput and the sum harvested energy in multiuser systems with fairness constraints.  Simulation results demonstrate that
substantial performance gains can be achieved by the proposed optimization framework compared to existing suboptimal scheduling algorithms from the literature.

\end{abstract}

\renewcommand{\baselinestretch}{0.92}
\large\normalsize
\section{Introduction}
\label{sec:intro}

Over the past decades, battery-powered devices have been deployed in many wireless communication networks. However, since batteries have limited energy storage capacity and their replacement can be costly or even infeasible, harvesting energy from the environment provides a viable solution for prolonging the network lifetime. Although conventional natural energy resources, such as solar and wind energy, are perpetual, they are   weather-dependent and location-dependent, which may not suitable for mobile communication devices. Alternatively, background radio frequency (RF) signals from ambient transmitters are also an abundant source of energy for energy harvesting (EH). Unlike the natural energy sources, RF energy is weather-independent  and can be available on demand. Nowadays, EH circuits are able to harvest microwatt to milliwatt of power over the range of several meters for a transmit power of $1$ Watt and a carrier frequency less than $1$ GHz  \cite{Powercast}. Thus, RF energy can be a viable energy source for devices with low-power consumption, e.g. wireless sensors \cite{Krikidis2014,Ding2014,JR:SWIPT_mag}. Moreover, RF EH provides the possibility for simultaneous wireless information and power transfer (SWIPT) since RF signals carry both information and energy \cite{Varshney2010,Grover2008}.

The integration of RF EH into communication systems introduces a paradigm shift in system and resource allocation algorithm design. A fundamental tradeoff between information and energy transfer rates was studied in \cite{Varshney2010, Grover2008}. However, current practical RF EH circuits are not yet able to harvest energy from an RF signal which was already used for information decoding (ID) \cite{Zhou2013}. To facilitate simultaneous ID and EH, a power splitting receiver was proposed in \cite{Zhou2013} and \cite{Zhang2013}. The energy efficiency of a communication system with power splitting receivers was investigated  in \cite{JR:WIPT_fullpaper}.
 In addition, a simple time-switching receiver has been proposed which switches between ID and EH in time. Furthermore, multiuser multiple input single output SWIPT systems were studied in \cite{Xu2013}, where beamformers were optimized for maximization of the sum harvested energy under minimum required signal-to-interference-plus-noise ratio constraints for multiple ID receivers. In \cite{JR:energy_beamforming}, the optimal energy transfer downlink duration was optimized to  maximize the uplink average information transmission rate. In \cite{vicky_SWIPT_2014}\nocite{JR:MOOP_SWIPT,JR:Kwan_SEC_DAS}-\hspace*{-1mm}\cite{kwan_SWIPT_2014},  beamforming design was studied for  secure SWIPT networks with different system configurations. In \cite{Ju2014}, a multiuser time-division-multiple-access system with energy transfer in the downlink (DL) and information transfer in the uplink was studied. The authors proposed a protocol for sum-throughput maximization and enhanced it by fair rate allocation among users with different channel conditions.  Nevertheless, multiuser scheduling, which exploits multiuser diversity for improving the system performance of multiuser systems, has not
been considered in  \cite{Varshney2010}-\hspace*{-1mm}\cite{Ju2014}. Recently,  simple suboptimal order-based schemes were proposed to balance the tradeoff between the users' ergodic achievable rates and their average amounts of harvested energy in \cite{Morsi2014}.  However, the  scheduling schemes proposed in \cite{Morsi2014} are unable to guarantee quality of service with respect to the minimum energy transfer.  In fact, optimal multiuser scheduling schemes that guarantee a long-term minimum harvested energy for SWIPT systems have not been considered in the literature  so far.

Motivated by the above observations,  we study  optimal scheduling schemes for long-term optimization which control the rate-energy (R-E) tradeoff under the consideration of  proportional fairness and equal throughput fairness.

\section{System Model}
\begin{figure}\centering\vspace*{-3mm}
\includegraphics[width= 2.5 in]{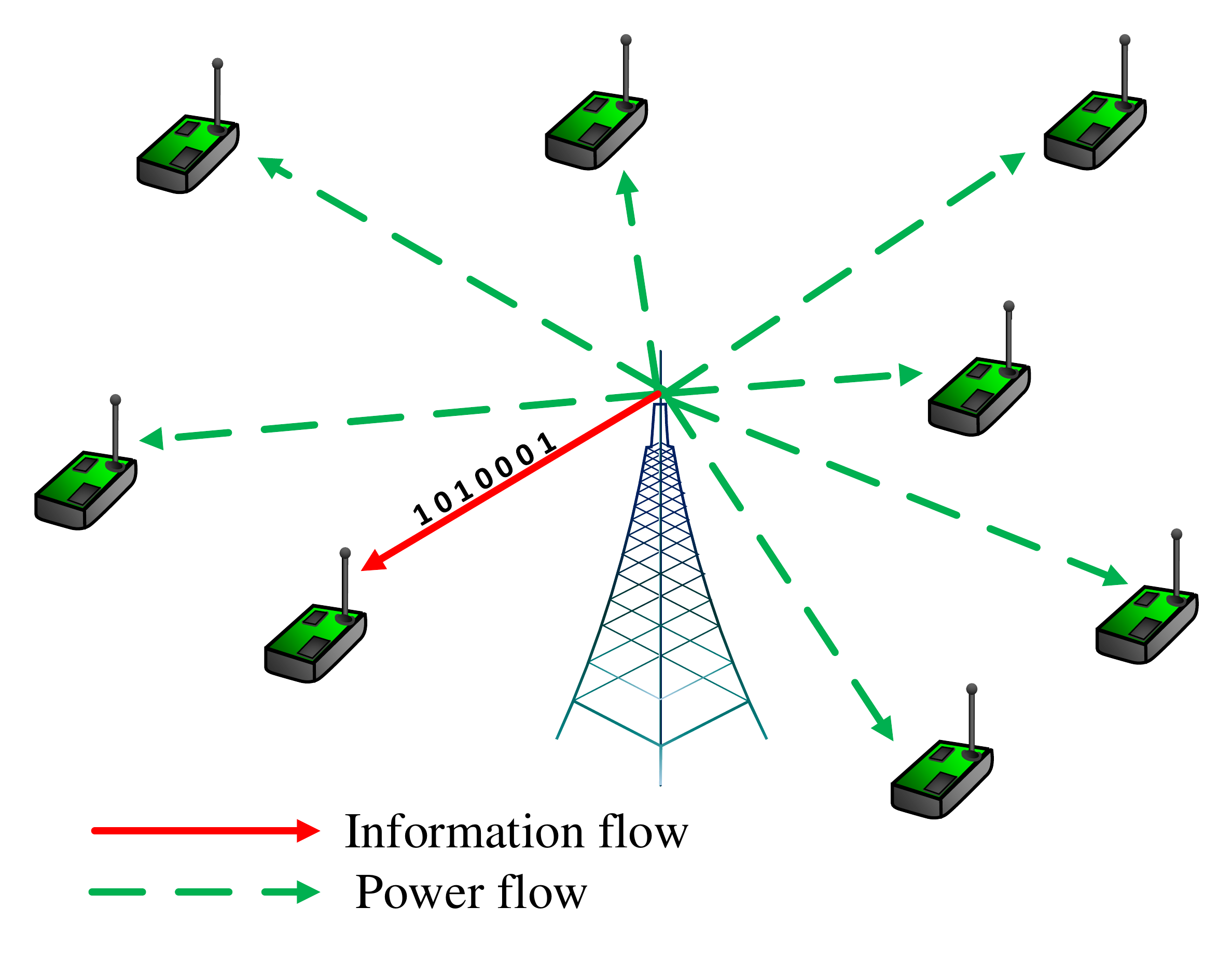}
\caption{A multiuser system with SWIPT for $N=8$ time-switching user terminals (UTs).}\vspace*{-3mm}
\label{fig:Fig4}

\end{figure}

We consider a SWIPT system that consists of one access point (AP) with a fixed power supply and $N$ battery-powered user terminals (UTs), see Fig. \ref{fig:Fig4}. The AP and the UTs are equipped with single antennas. Besides, we adopt time-switching receivers at the UTs \cite{Krikidis2014}  to ensure low hardware
complexity.
We study the user scheduling for DL transmission. We assume that the transmission is divided into $T$ time slots and  in each time slot perfect CSI is available at the AP. Also, the data buffer for the users at the AP is always full such that enough data packets are available for transmission for every scheduled UT. In each time slot, the AP schedules one user for ID, while the remaining users opportunistically harvest energy\footnote{We consider a unit-length time slot, hence the terms ``power" and ``energy" can be used interchangeably.} from the received signal.  We assume block fading channels. In particular, the channels remain constant during a time slot and change independently over different time slots. Besides, the users are physically separated from one another such that they experience independent fading.

 Furthermore, we adopt the  EH receiver model  from \cite{Zhou2013}. The RF energy harvested by user $n \in \{ 1, \ldots, N\}$ in time slot $i\in\{1,\ldots,T\}$ is given by
\begin{equation}
Q_n(i) = \xi_n P h_n(i),
\end{equation}
where $P$ is the constant AP transmit power, $0 \leq \xi_n \leq 1$ is the RF-to-direct-current (DC) conversion efficiency of the EH receiver of user $n$,  and  $h_n(i)$ is the channel power gain between the AP and user $n$ in time slot $i$.

\section{Optimal Multiuser Scheduling}
In the following, we propose three optimal multiuser scheduling schemes that control the R-E tradeoff under different fairness considerations.
\subsection{Optimal Maximum Throughput (MT) Scheme}
First, we consider a scheduling scheme which maximizes the average sum rate subject to a constraint on the minimum required average aggregate harvested energy.  We note that this scheme aims to reveal the best system performance, and  fairness in resource allocation for UTs is not considered. To facilitate the following presentation, we introduce the user selection variables $q_n (i)$, where $i \in \{1, 2, \ldots T\}$ and $n \in \{ 1, \ldots , N \}$. In time slot $i$, if user $n$ is scheduled to perform ID, $q_{n} (i) = 1$, whereas $q_{\bar{n}} (i) = 0, \forall  \bar{n} \neq n$, i.e., all the remaining idle users  harvest energy from the transmitted signal. Now, we formulate the MT optimization problem as follows.

\begin{Prob}{Maximum Throughput Optimization:}\end{Prob}\vspace*{-3mm}
\begin{equation}
\begin{aligned}
& \underset{q_n (i), \forall n, i}{\mathrm{maximize}}
& & \bar{R}_{\mathrm {sum}} \\
& \text{subject to}
& & \text{C1:} \hspace{7pt} \sum^N_{n=1} q_n (i) = 1, \forall i, \\
&&& \text{C2:} \hspace{7pt} q_n (i) \in \{ 0, 1 \}, \forall n, i, \\
&&& \text{C3:} \hspace{7pt} \bar{Q}_{\mathrm{sum}} \geq Q_{\mathrm{req}},
\end{aligned}
\label{eq:MTbin}
\end{equation}
where
\begin{alignat}{2}
\bar{R}_{\mathrm {sum}} &= \lim_{T \to \infty} \frac{1}{T} \sum^{T}_{i=1}\sum^{N}_{n=1} q_n (i) C_n (i), \\
 \bar{Q}_{\mathrm {sum}} &= \lim_{T \to \infty} \frac{1}{T} \sum^{T}_{i=1}\sum^{N}_{n=1} (1-q_n (i)) Q_n (i),\, \mbox{and} \\
 C_n(i) &= \log_2 \left(1 + \frac{P h_n(i)}{\sigma_n^2}\right).
\end{alignat}
Here, $\sigma_n^2$ is the additive white Gaussian noise power at UT $n$. In the considered problem, we focus on the long-term system performance for $T \rightarrow \infty$. Constraints C1 and C2 ensure that in each time slot only one user is selected to receive information. C3 ensures that the average amount of harvested energy $\bar{Q}_{\mathrm {sum}}$ is no less than the minimum required amount $Q_{\mathrm {req}}$. Since the user selection variables $q_n (i), \forall n, i$, are binary, problem \eqref{eq:MTbin} is non-convex. In order to handle the non-convexity, we adopt the time-sharing relaxation. In particular, we relax the binary constraint C2 such that $q_n (i)$ is a continuous value between zero and one.  Then, the relaxed version of problem  (\ref{eq:MTbin}) can be written  in minimization form as:
\begin{equation}
\begin{aligned}
& \underset{q_n (i), \forall n, i}{\mathrm{minimize}}
& & -\bar{R}_{\mathrm {sum}} \\
& \text{subject to}
& & \text{C1, C3}, \\
&&& \widetilde{\text{C2}}: \hspace{7pt}  0 \leq q_n (i) \leq 1, \forall n, i.
\end{aligned}
\label{eq:MTrel}
\end{equation}
Now,  we introduce the following theorem that reveals the tightness of the binary constraint relaxation.
\begin{Thm}Problems (\ref{eq:MTbin}) and (\ref{eq:MTrel}) are equivalent\footnote{Here, ``equivalent" means that both problems share the same optimal $q_n(i)$. } with probability one, when $h_n (i), \forall n, i$ are independent and continuously distributed. In particular, the constraint relaxation of C2 is tight, i.e.,\end{Thm}\vspace*{-4mm}
\begin{equation}
\mbox{C2} \Leftrightarrow \widetilde{\text{C2}}: 0 \leq q_n (i) \leq 1, \forall n, i. \label{eq:Theorem}
\end{equation}

\begin{proof}
Theorem 1 will be proved in the following based on the optimal solution of (\ref{eq:MTrel}).
\end{proof}
In other words, we can solve  (\ref{eq:MTbin}) via solving (\ref{eq:MTrel}). It can be verified that the relaxed problem is convex with respect to the relaxed optimization variables and satisfies the Slater's constraint qualification.
Therefore, strong duality holds and the optimal solution of \eqref{eq:MTrel} is equal to the optimal solution of its dual problem \cite{Boyd2004}. Thus, we solve \eqref{eq:MTrel} via the dual problem. To this end, we first define the Lagrangian function for the above optimization problem as
\begin{align}
&  L(q_n(i), \lambda (i), \alpha_n (i), \beta_n(i), \nu ) = - \bar{R}_{\mathrm {sum}} \notag \\
& + \sum^T_{i=1} \lambda (i) \left( \sum^N_{n=1} q_n(i) -1 \right)
+ \sum^T_{i=1} \sum^N_{n=1} \alpha_n(i) \left( q_n(i) -1 \right) \notag \\
& - \sum^T_{i=1}\sum^N_{n=1} \beta_n(i) q_n(i) + \nu \left( Q_{\mathrm {req}} - \bar Q_{\mathrm {sum}} \right),
\label{eq:MTLagr}
\end{align}where $\lambda (i), \{ \alpha_n (i) \geq 0, \beta_n(i) \geq 0 \}$, and $\nu \geq 0$ are the Lagrange multipliers corresponding to constraints C1, $\widetilde{\text{C2}}$, and C3, respectively. Thus, the dual problem of (\ref{eq:MTrel}) is given by
\begin{eqnarray}
\underset{\alpha_n (i), \beta_n(i)\ge 0, \lambda (i)}{\mathrm{maximize}} \underset{q_n(i)}{\mino}\,  L(q_n(i), \lambda (i), \alpha_n (i), \beta_n(i), \nu ).\hspace{-5pt}
\end{eqnarray}
In order to determine the optimal user selection policy, we apply standard convex optimization techniques and the Karush-Kuhn-Tucker (KKT) conditions. Thereby, we differentiate  the Lagrangian in \eqref{eq:MTLagr} with respect to $q_n (i)$ and set it to zero which yields:
\begin{equation}
\frac{\partial L (\ldots)}{\partial q_n(i)} = \lambda(i) + \alpha_n(i) - \beta_n(i) +
\nu \frac{Q_n(i)}{T}  - \frac{C_n(i)}{T}  = 0, \,  \forall n,i.\label{eq:StatCond}
\end{equation}
We define $n^*$ as the optimal user selection index for ID at time slot $i$, i.e., $q_{n^*} (i) = 1$ and $q_n(i) = 0, \hspace{3pt} \forall n \neq n^*$. From the complementary slackness condition, we obtain $\alpha_n(i) = 0, \hspace{3pt} \forall n \neq n^*$ and $\beta_{n^*}(i) = 0$.
Now, we denote the optimal dual variable for constraint C3 as  $\nu^*$ and substitute it into \eqref{eq:StatCond}. Then, the  selection metric for UT $n$ is given as
\begin{subequations}
\begin{alignat}{3}
&\Lambda_n^{\mathrm{MT}}(i) = T \left( \lambda (i)\hspace*{-0.5mm} + \hspace*{-0.5mm}\alpha_n(i)- \beta_n(i) \right) = C_n(i) \hspace*{-0.5mm}-\hspace*{-0.5mm} \nu^* Q_n(i). \label{eq:MetricGen}\\
& \text{Hence, the selection metric for the scheduled UT is} \notag \\
& \Lambda^{\mathrm{MT}}_{n^*}(i) = T \left( \lambda (i) + \alpha_{n^*}(i) \right) = C_{n^*}(i) - \nu^* Q_{n^*}(i), \label{eq:MetricNStar} \\
& \text{and the selection metric for the remaining UTs is} \notag \\
& \Lambda^{\mathrm{MT}}_n(i) = T \left( \lambda (i) - \beta_n(i) \right) = C_n(i) - \nu^* Q_n(i), \hspace{5pt} \forall n \neq n^* . \label{eq:MetricN}
\end{alignat}
\label{eq:MetricAll} \end{subequations}

\hspace{-15pt}Subtracting \eqref{eq:MetricN} from \eqref{eq:MetricNStar} yields
\begin{equation}
 \Lambda^{\mathrm{MT}}_{n^*}(i) - \Lambda^{\mathrm{MT}}_n(i) = T \left( \alpha_{n^*}(i) + \beta_n (i) \right).
\label{eq:Diff} \end{equation}
Since $\alpha_{n^*} (i) + \beta_n(i) \geq 0$ from the dual feasibility conditions, we obtain  $\Lambda^{\mathrm{MT}}_{n^*}(i) \geq \Lambda^{\mathrm{MT}}_n(i), \hspace{3pt} \forall n \neq n^*$. Furthermore, $\Lambda^{\mathrm{MT}}_n(i) \hspace{3pt} \forall n$, are continuous random variables, therefore $\Pr \{ \Lambda^{\mathrm{MT}}_{n^*}(i) = \Lambda^{\mathrm{MT}}_n(i) \}= 0, \hspace{3pt} \forall i$, where $\Pr \{ \cdot \}$ denotes the probability of an event. Thus, $\Lambda^{\mathrm{MT}}_{n^*}(i) > \Lambda^{\mathrm{MT}}_n(i), \hspace{3pt} \forall n \neq n^*$ and the optimal selection criterion for the MT scheme in time slot $i$ reduces to
\begin{equation}
\label{eq:MTSelRule} q_{n}(i)=
 \left\{ \begin{array}{rl}
 1 &\mbox{if $\Lambda^{\mathrm{MT}}_n(i)=\underset{t\in\{1,\ldots,N\}}{\max} \,\,\Lambda^{\mathrm{MT}}_t(i)$}\\
 0 &\mbox{otherwise}
       \end{array} \right..
\end{equation}
In other words,  the solution of the relaxed problem
is itself of the Boolean type. Therefore, the adopted binary relaxation is tight. Besides, $\nu^*$ depends only on the statistics of the channels. Hence, it can be calculated offline, e.g. using the gradient method, and then used for online scheduling as long as the channel statistics remain unchanged. We emphasize that although the original problem in \eqref{eq:MTrel} considers infinite number of time slots and long-term averages for the sum rate and the total harvested energy, the optimal scheduling rule in \eqref{eq:MTSelRule} depends only on the current time slot, i.e., online scheduling is optimal.

\subsection{Optimal Proportional Fair (PF) Scheme}
In the MT scheme, UTs with weak channel conditions may be deprived from gaining access to the channel which leads to user starvation. In order to strike a balance between system throughput and fairness, we introduce proportional fairness  into our scheduler, which aims to provide each UT with a performance proportional to its channel conditions. This is achieved by allowing all UTs to access the channel with  equal chances. In this case, the optimization problem with the relaxed binary constraint on the user selection variables is formulated as:
\begin{Prob}{Optimal Proportional Fair Optimization:}\end{Prob}\vspace*{-4mm}
\begin{equation}
\begin{aligned}
& \underset{q_n (i), \forall n, i}{\mathrm{minimize}}
& & -\bar{R}_{\mathrm {sum}} \\
& \text{subject to}
& & \text{C1, $\widetilde{\text{C2}}$, C3}, \\
&&& \text{C4:} \hspace{7pt}  \frac{1}{T} \sum^T_{i=1} q_n (i) - \frac{1}{N} = 0,\forall n,
\end{aligned}
\label{eq:PFrel}
\end{equation}
where C4 specifies that each UT has to access the channel for $\frac{T}{N}$ number of time slots. For the tightness of the binary relaxation, please refer to  Theorem 1.

Now, we solve (\ref{eq:PFrel}) via convex optimization techniques by following a similar approach as in the previous section. The Lagrangian function for problem \eqref{eq:PFrel} is given by
\begin{align}
& L(q_n(i), \lambda (i), \alpha_n (i), \beta_n(i), \nu, \gamma_n ) = - \bar{R}_{\mathrm {sum}} \notag \\
& + \sum^T_{i=1} \lambda (i) \left( \sum^N_{n=1} q_n(i) -1 \right) + \sum^T_{i=1} \sum^N_{n=1} \alpha_n(i) \left( q_n(i) -1 \right) \notag \\
& - \sum^T_{i=1}\sum^N_{n=1} \beta_n(i) q_n(i) + \nu \left( Q_{\mathrm {req}} - \bar Q_{\mathrm {sum}} \right) \notag \\
& + \sum^N_{n=1} \gamma_n \left( \frac{1}{T} \sum^T_{i=1} q_n (i) - \frac{1}{N} \right),\label{eq:PFLagr}
\end{align}
where $\lambda (i), \{\alpha_n (i) \geq 0, \beta_n(i) \geq 0\}, \nu \geq 0,$ and $\gamma_n$ are the Lagrange multipliers corresponding to constraints C1, $\widetilde{\text{C2}}$, C3, and C4, respectively.
By using the KKT conditions,  we obtain the following UT selection metric:
\begin{equation}\label{eqn:UT_metric}
\Lambda^{\mathrm{PF}}_n(i)=C_n (i) - \nu^* Q_n (i) - \gamma^*_n,
\end{equation}
where the optimal Lagrange multipliers $\gamma^*_n$ ensure that each user accesses the channel on average an equal number of times.
Thus, the optimal selection criterion for the PF scheme is
\begin{equation}
\label{eq:PFSelRule} q_{n}(i)\hspace*{-0.5mm}=\hspace*{-0.5mm}
 \left\{ \begin{array}{rl}
 \hspace*{-2mm}1 &\hspace*{-2mm}\mbox{if $\Lambda^{\mathrm{PF}}_n(i)=\hspace*{-2mm}\underset{t\in\{1,\ldots,N\}}{\max} \,\,\Lambda^{\mathrm{PF}}_t(i)$}  \\
 \hspace*{-2mm}0 &\hspace*{-2mm}\mbox{otherwise}
       \end{array} \right..
\end{equation}
We note that the optimal PF scheduling rule  is similar to the MT scheduling rule in \eqref{eq:MTSelRule}, but the PF seelction metric in (\ref{eqn:UT_metric}) contains an additional term $\gamma^*_n$ that provides proportional fairness. Also, $\nu^*$ and $\gamma^*_n$ can be calculated offline using the gradient method.

\subsection{Optimal Equal Throughput (ET) Scheme}
Although the PF scheduler enables equal channel access probability for all UTs, it does not provide any guaranteed minimum data rate to them.  On the contrary, the ET criterion is more fair from the users' prospective compared to the PF criterion, as all the UTs achieve the same average throughput asymptotically for $T\rightarrow\infty$. Therefore, in this section, we design a scheduler which achieves ET fairness.  Thus, the objective is to maximize the minimum average achievable rates among all the UTs, i.e., maximize $\min\limits_{n} \bar{C}_n$ where $\bar{C}_n=\lim_{T \to \infty} \frac{1}{T} \sum^{T}_{i=1} q_n (i) C_n (i)$. Using Theorem 1, we formulate our equivalent convex optimization problem in its hypograph form.\newpage
\begin{Prob}{Optimal Equal Throughput Optimization:}\end{Prob}\vspace*{-4mm}
\begin{equation}
\begin{aligned}
& \underset{r, q_n (i), \forall n, i}{\mathrm{minimize}}
& & - r \\
& \text{subject to}
& & \text{C1, $\widetilde{\text{C2}}$, C3}, \\
&&& \text{C5:} \hspace{7pt}  r - \bar{C}_n \leq 0, \hspace{5pt} \forall n,
\end{aligned}
\label{eq:ETrel}
\end{equation}
where $r$ is an auxiliary variable.  The Lagrangian function for problem in \eqref{eq:ETrel} is given by
\begin{align}
& L(q_n(i), \lambda (i), \alpha_n (i), \beta_n(i), \nu, \theta_n ) = -r  \notag \\
&+ \sum^T_{i=1} \lambda (i) \left( \sum^N_{n=1} q_n(i) -1 \right) + \sum^T_{i=1} \sum^N_{n=1} \alpha_n(i) \left( q_n(i) -1 \right) \notag \\
& - \sum^T_{i=1}\sum^N_{n=1} \beta_n(i) q_n(i)\hspace*{-0.5mm} +\hspace*{-0.5mm} \nu \left( Q_{\mathrm {req}}\hspace*{-0.5mm} - \hspace*{-0.5mm} \bar Q_{\mathrm {sum}} \right)\hspace*{-0.5mm}+ \hspace*{-0.5mm}\sum^N_{n=1} \theta_n \left( r- \bar{C}_n \right), \label{eq:ETLagr}
\end{align}
where $\lambda (i), \{ \alpha_n (i) \geq 0, \beta_n(i) \geq 0 \}, \nu\geq 0$, and $\theta_n\geq 0$ are the Lagrange multipliers corresponding to constraints C1, $\widetilde{\text{C2}}$, C3, and C5, respectively.
By using the KKT conditions,  we obtain the UT selection metric for ET scheduling:
\begin{equation}
\Lambda^{\mathrm{ET}}_n(i)=\theta^*_n C_n(i) - \nu^* Q_n(i),
\end{equation}
where the optimal Lagrange multipliers $\theta^*_n$ ensure that all users have ET. Thus, the optimal selection criterion for the ET scheme is given by
\begin{equation}
\label{eq:ETSelRule} q_{n}(i)\hspace*{-0.5mm}=\hspace*{-0.5mm}
 \left\{ \begin{array}{rl}
 \hspace*{-2mm}1 &\hspace*{-2mm}\mbox{if $\Lambda^{\mathrm{ET}}_n(i)=\hspace*{-2mm}\underset{t\in\{1,\ldots,N\}}{\max} \,\,\Lambda^{\mathrm{ET}}_t(i)$} \\
 \hspace*{-2mm}0 &\hspace*{-2mm}\mbox{otherwise}
       \end{array} \right. .
\end{equation}
Again, the gradient method can be used to obtain the optimal values for $\nu^*$ and $\theta_n^*$ offline by utilizing the channel statistics.
\begin{Remark}
We note that the above considered problems can be formulated as Markov Decision Process (MPD) or solved via Lyapunov optimization approach, please refer to \cite{book:MDP} for details.
\end{Remark}
\section{Simulation Results}
In this section, we evaluate the performance of the proposed scheduling schemes using simulations. The important simulation parameters are summarized in Table 1. We adopt the path loss model from \cite{Rappaport} and the UTs are randomly and uniformly distributed between the reference distance and maximum service distance. For comparison,  we also show the performance of the following suboptimal scheduling schemes from \cite{Morsi2014}:

\begin{table}\caption{Simulation parameters.}

\begin{center}
\renewcommand{\arraystretch}{1.2}
\begin{tabular}{ll}
\hline
\textbf{Parameter}    & \textbf{Value} \\
\hline
AP transmit power $P$ & $40$ dBm \\
Noise power $\sigma^2_n$ & $-62$ dBm \\
RF-to-DC conversion efficiency $\xi_n$ & $0.5$ \\
Path loss exponent & $3.6$ \\
Maximum service distance &  $100$ m \\
Reference distance & $2$ m \\
Antenna gain of AP and UTs & $10$ dBi \& $2$ dBi \\
Carrier center frequency & $915$ MHz \\
Bandwidth & $200$ kHz \\
Fading channel & Rayleigh\\
\hline
\end{tabular}
\end{center}
\vspace*{-0.3cm}

\label{tab:res}\vspace*{-0.3cm}
\end{table}

\begin{figure}[t]  \centering
\includegraphics[width= 3.5 in]{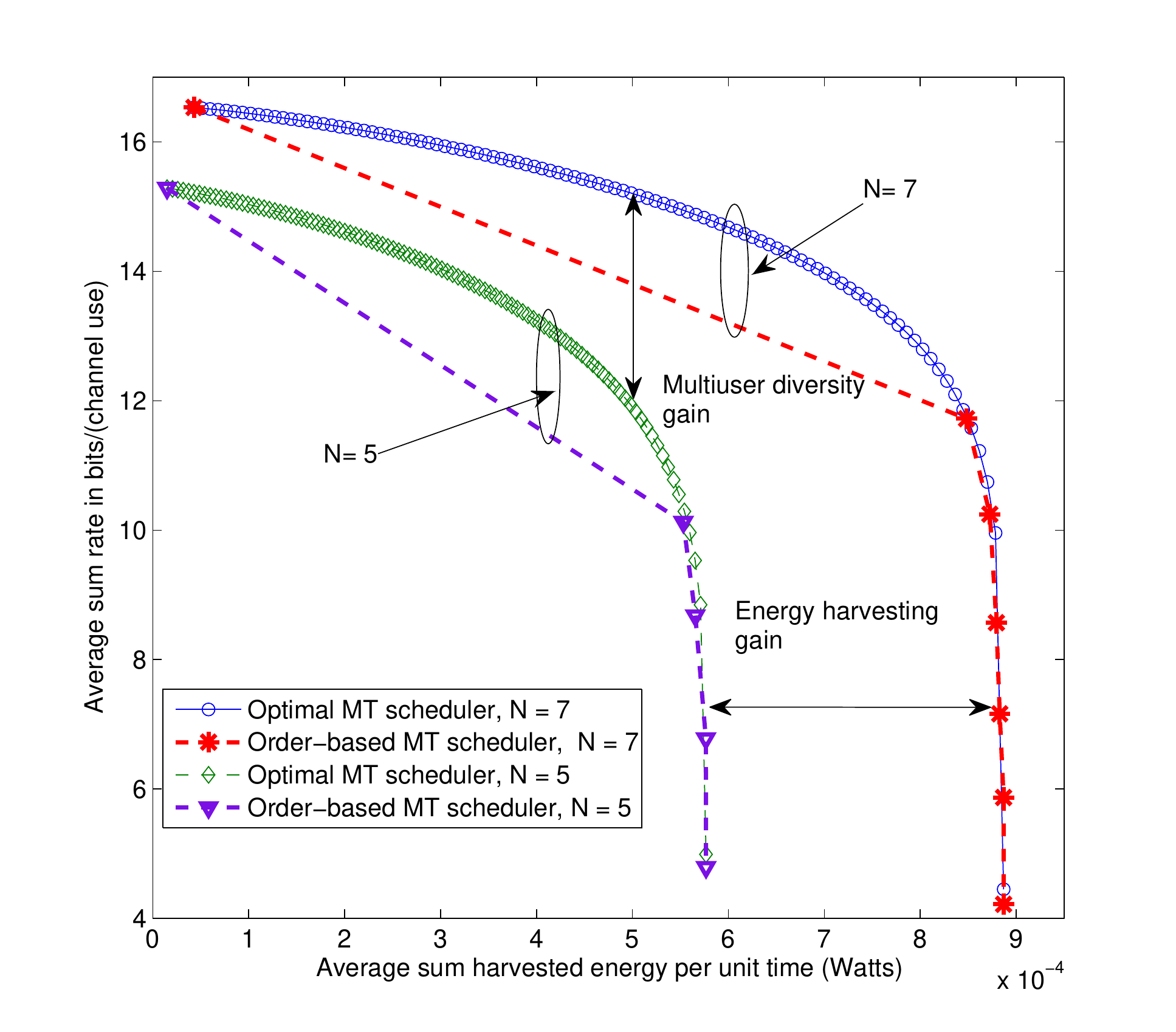}
\vspace*{-0.9cm}
\caption{Average sum rate versus average sum harvested energy of the MT schemes for different numbers of UTs.}\vspace*{-0.2cm}
\label{fig:MT2}
\end{figure}

\begin{enumerate}
\item Order-based MT scheduler: The scheduling rule is $n^* (i) = \argord \limits_{n\in\{1,\ldots,N\}} h_n (i),$ where $\argord$ is defined as the argument of a certain selection order $j \in \{1, \ldots, N \}$. In other words, the user whose channel power gain $h_n(i)$ has order $j$ is scheduled for ID.
\item Order-based PF scheduler: The scheduling rule is $n^* (i) = \argord\limits_{n\in\{1,\ldots,N\}} \frac{h_n (i)}{\Omega_n},$
where $\Omega_n$ denotes the mean channel power gain of UT $n$.
\item Order-based ET scheduler: The scheduling rule is $n^* (i) =\arg\min\limits_{{O_{\Un_n}\in \Sa}} r_n(i-1),$ where $O_{\Un_n}\in\{1,\ldots,N\}$ is the order of the  instantaneous normalized signal-to-noise-ratio of user $n$, $\Sa$ is a predefined set of orders, where only the users set $O_{\Un_n}\in\{1,\ldots,N\}$ fall into $\Sa$ are eligible for being scheduled, and $r_n(i-1)$ is the throughput of user $n$ averaged over all previous time slots up to time slot $i-1$.
\end{enumerate}
Fig. \ref{fig:MT2} shows the average sum rate (bits/(channel use)) versus the average sum harvested energy (Watts) of the MT schemes for different numbers of users. We note that the suboptimal order-based scheme can only achieve discrete points on the R-E curves, corresponding to the selection orders $j\in\{1,\ldots,N\}$.   On the contrary, the proposed optimal MT scheduling scheme can achieve any feasible point on the R-E curve, which provides a higher flexibility for the system designer to strike a balance between average sum rate and average harvested energy. Besides, as expected,  the average system sum rate  increases with the number of UTs as the proposed scheme is able to exploit multiuser diversity. Furthermore, the average sum harvested energy also increases with the number of UTs since more idle users participate in energy harvesting in any given time slot.

\begin{figure}[t]  \centering\vspace*{-0.9cm}
\includegraphics[width= 3.5 in]{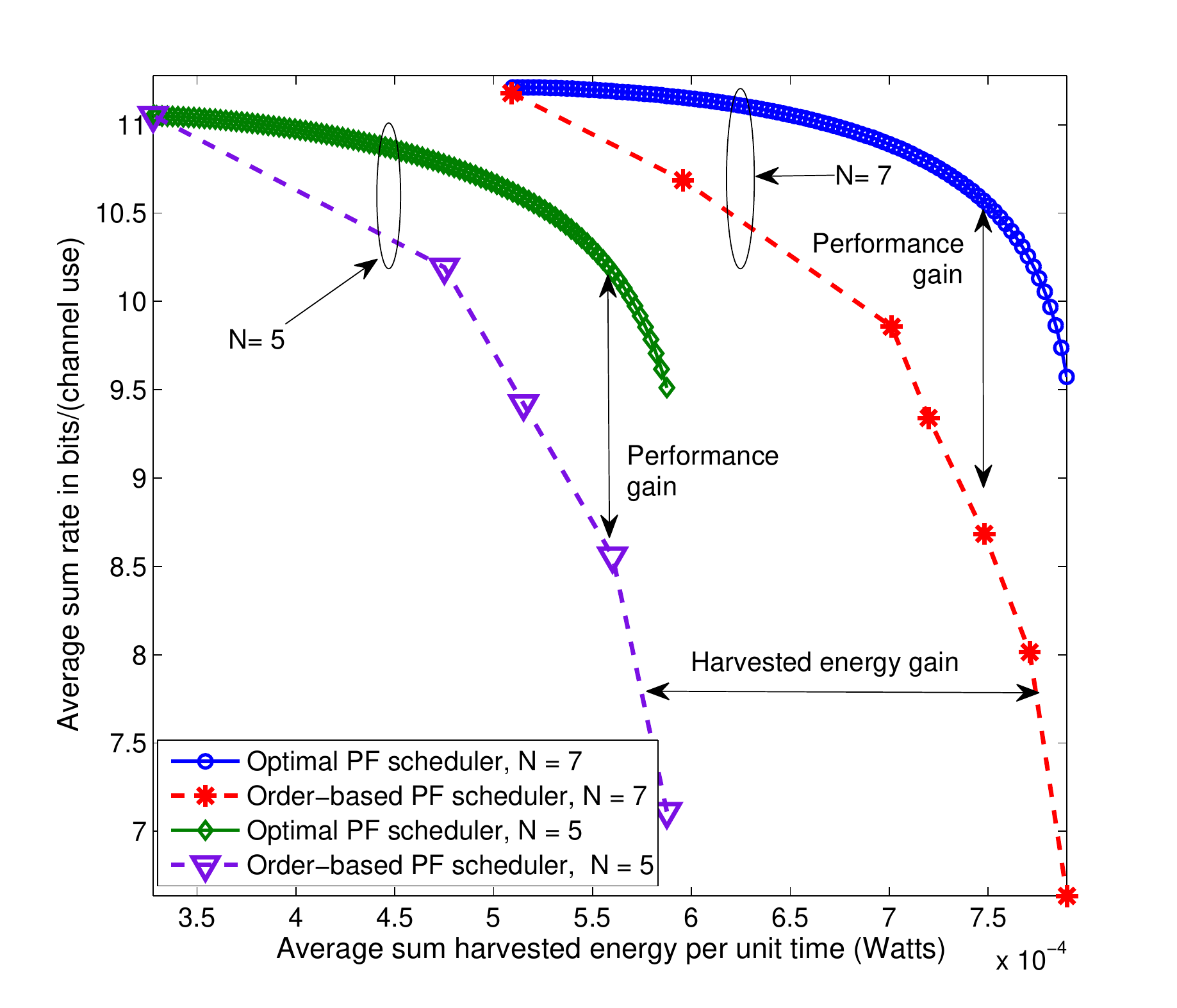}
\vspace*{-0.9cm}
\caption{Average sum rate versus average sum harvested energy of the PF schemes for different numbers of UTs.}
\label{fig:PF2}
\end{figure}

\begin{figure}[t]
  \centering\vspace*{-0.9cm}
\includegraphics[width=3.5 in]{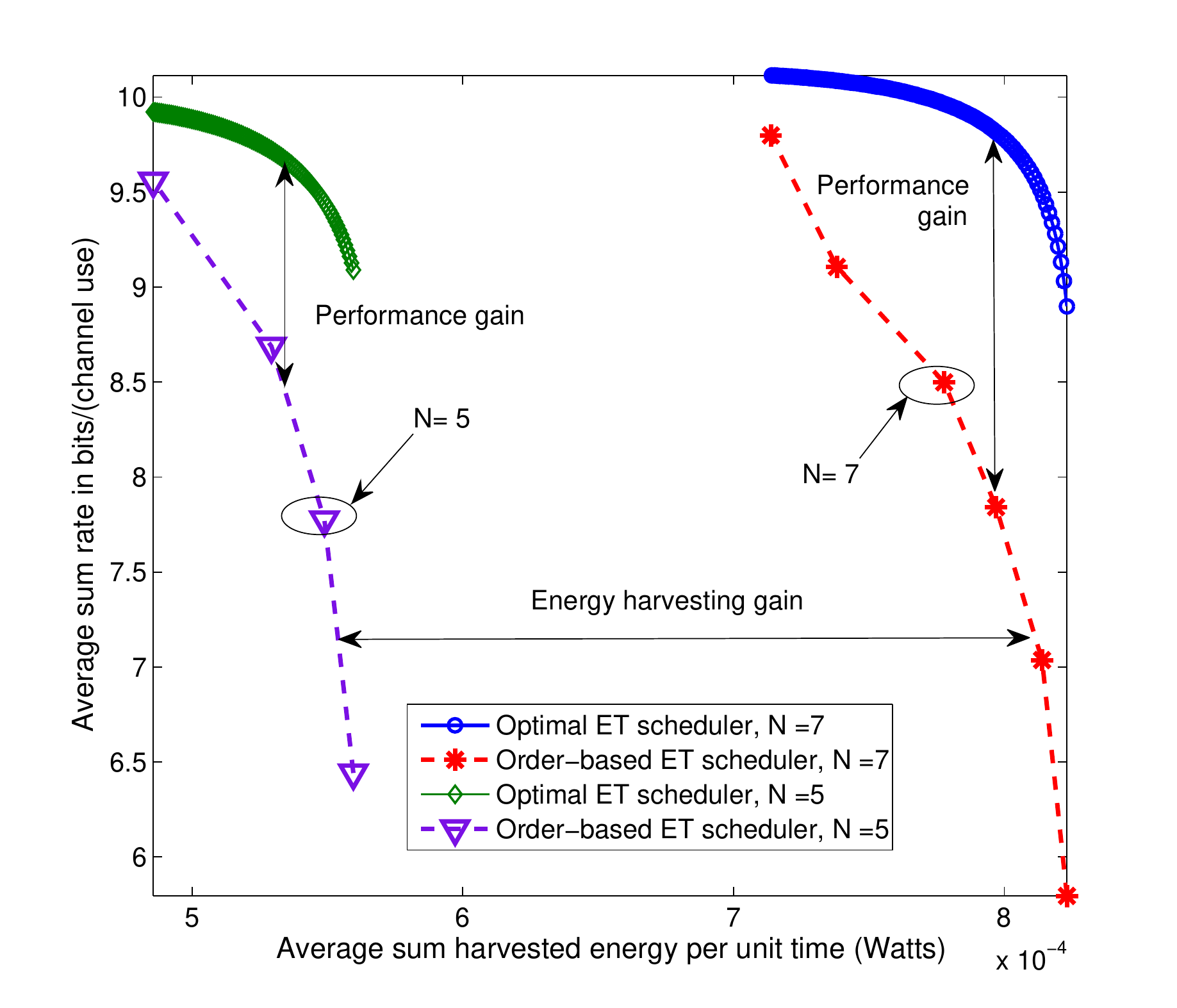}
\vspace*{-0.9cm}
\caption{Average sum rate versus average sum harvested energy of the ET schemes for different numbers of UTs.}
\label{fig:ET2}

\end{figure}

Fig. \ref{fig:PF2} and Fig. \ref{fig:ET2} depict the average sum rate (bits/(channel use)) versus the average sum harvested energy (Watts) for the PF and ET schemes, respectively. It can be seen that the feasible R-E region of all schemes decreases compared to the MT scheduler in Fig. \ref{fig:MT2}. This is because both the PF and the ET schedulers take fairness into account in the resource allocation and, as a result, cannot fully exploit the multiuser diversity for improving the average system sum rate. On the other hand, it can be seen that our proposed optimal schemes provide a substantial average sum rate  gain compared to the corresponding suboptimal order-based schemes, especially for a high amount of average harvested energy in the system.  In fact, the proposed optimization framework provides more degrees of freedom across different time slots in resource allocation compared to the suboptimal scheduling schemes.  This allows the system to exploit the multiuser diversity to some extent for resource allocation even if fairness is taken into consideration.

\section{Conclusion}
In this paper, we have proposed optimal multiuser scheduling schemes for SWIPT systems considering different notions of fairness in resource allocation. The designed schemes enable the control of the tradeoff between the average sum rate and the average amount of sum harvested energy. Our results reveal that for the maximization of the system sum rate with or without fairness constraints, the optimal scheduling algorithm  requires only causal  instantaneous and statistical channel knowledge. Simulation results revealed that
substantial performance gains can be achieved by the proposed optimization framework compared to existing suboptimal scheduling schemes.
\bibliographystyle{IEEEbib}
{
\footnotesize\bibliography{strings,refs}}

\end{document}